\documentclass[preprint,10pt,aip,numerical,jcp]{revtex4-2}

\usepackage[english]{babel}	
\usepackage[T1]{fontenc}
\usepackage[utf8]{inputenc}	

\usepackage{mathptmx} 
\usepackage{mathtools,amssymb,nccmath}	
\usepackage{amsmath,amsthm,amscd}
\usepackage{amsbsy}
\usepackage{amsfonts}
\usepackage{mathrsfs}
\usepackage{braket}
\usepackage{bm}          	
\usepackage{url}		
\usepackage{breakurl}
\usepackage[breaklinks]{hyperref}
\usepackage{hyperref} 
\usepackage{multirow}
\usepackage{color}

\usepackage{blindtext}

\makeatletter
\newcommand{\optionaldesc}[2]{%
  \phantomsection
  #1\protected@edef\@currentlabel{#1}\label{#2}%
}
\makeatother

\draft 

\makeatletter
\newcommand*{\rom}[1]{\expandafter\@slowromancap\romannumeral #1@} 
\makeatother
\allowdisplaybreaks 

\begin{document}


\title{Density dependent embedding potentials for piecewise exact densities}

\author{Tomasz A. Wesolowski}
\altaffiliation{tomasz.wesolowski@unige.ch.}
\affiliation{Department of Physical Chemistry, University of Geneva, Quai Ernest-Ansermet 30,\\
CH-1211 Gen\`eve 4, Switzerland}

\date{\today}

\begin{abstract}\label{abstract}
%
%
%
Frozen Density Embedding Theory (FDET) [Wesolowski  {\it Phys. Rev. A} {\bf 77}, 012504 (2008)] provides the interpretation
of the eigenvalue equations for an embedded $N'$-electron wavefunction, in which the embedding operator is  multiplicative,   as the Euler-Lagrange equation corresponding to the constrained minimisation of  the Hohenberg-Kohn energy functional. The constraint is given by a non-negative function integrating to an integer $N-N'>0$ with $N$ being the total number of electrons  in the whole system ($\min_{\rho\rightarrow\forall_{\mathbf{r}}\big(\rho({\mathbf r})\ge \rho_2({\mathbf r}\big)} E^{HK}_v[\rho]=E^{HK}_v[\rho_1^{o}+\rho_2]\ge E^{HK}[\rho_v^{o}]=E^o_v$).
The necessary condition that $\rho_1^{o}({\mathbf r})=\rho_v^{o}({\mathbf r})-\rho_2({\mathbf r})$ and consequently that $E^{HK}_v[\rho_1^{o}+\rho_2]=E^{HK}_v[\rho_v^{o}]=E^o_v$
is obviously that $\rho_2({\mathbf r})\le \rho_v^{o}({\mathbf r})$ on any measurable volume element.
In this work, we show that the equality $\rho_1^{o}({\mathbf r})=\rho_v^{o}({\mathbf r})-\rho_2({\mathbf r})$ imposes a stronger condition on $\rho_B$, namely, $\rho_2({\mathbf r})< \rho_v^{o}({\mathbf r})$.   
Properties of the exact solution of the FDET eigenvalue equations are derived if this condition is not met.  
The result is discussed in the context of subsystem approach to the Hohenberg-Kohn theorems, pseudopotential theory, and embedding potentials derived from inverted of Kohn-Sham equations.

\end{abstract}

\maketitle


%
%
%
\subsubsection*{Preliminaries: definitions, notation, statement of the problem}

This work concerns the following minimisation problem: 
\begin{eqnarray}
\min_{\rho\rightarrow\forall_{\mathbf{r}}\big(\rho({\mathbf r})\ge \rho_2({\mathbf r}\big)} 
E^{HK}_v[\rho]=
\min_{\rho_1\rightarrow N'}E^{HK}_v[\rho_1+\rho_2]
=E^{HK}_v[\rho_1^{o}+\rho_2]\ge E^o_v\label{eq:constrminim}
\end{eqnarray}   
 for the Hohenberg-Kohn functional $E_v^{HK}[\rho]$ in a system of $N$-electrons in a Coulombic external potential $v$, for such $\rho_2$ that  $\forall_{\mathbf {r})}\rho_2(\mathbf {r})\ge 0$, $\int \rho_2(\mathbf {r})d\mathbf{r}= N-N'$ with integer $N$.
 
Throughout this work, $\rho$ and $\rho_1$  denote non-specific functions on which the considered functionals depend.
Specific functions are indicated by additional subscripts or superscripts. 
All considerations are made for a given Coulombic external potential ($v$), total number of electrons ($N$), and the number of embedded electrons ($N'$). 
All formulas are given in atomic units.

Eq. \ref{eq:constrminim} provides thus the definition of $\rho_1^{o}[\rho_2]$. 
Square brackets in  $\rho_1^{o}[\rho_2]$ indicate  that $\rho_1^{o}$  is uniquely determined by  $\rho_2$ for a given external potential $v$.
Eq. \ref{eq:constrminim} provides also the definition of the functional $E^{\rho_2}_{v}[\rho_1]$:
\begin{eqnarray}
E^{\rho_2}_{v}[\rho_1]\equiv E_v^{HK}[\rho_1+\rho_2],\label{def:errho2v}
\end{eqnarray}
which for a given potential $v$ and the $N-N'$-electron density $\rho_2$. 
The domain of admissible functions for $E^{\rho_2}_{v}[\rho_1]$  comprises $N'$-electron densities. $\rho_1^{o}$ defined in Eq. \ref{eq:constrminim} is a minimiser of  $E^{\rho_2}_{v}[\rho_1]$ and not of the Hohenberg-Kohn functional.
Frozen-Density Embedding Theory (FDET)  \cite{Wesolowski2008} provides the formal framework to perform the above search for the minimiser of  $E^{\rho_2}_{v}[\rho_1]$ for given $v$ and $\rho_2$ by means of solving an eigenvalue problem for  a smaller number of electrons than $N$. 

%
\begin{eqnarray}
    \left( \hat{T}^{N'}+\hat{V}^{N'}+\hat{V}_{ee}^{N'}
    + \hat{V}_{emb}^{N'}
    \right) \Psi^{N'} = \lambda^{N'}\cdot\Psi^{N'}, \label{eq:emb_SE} 
 \end{eqnarray}
where 
%
%
 \begin{eqnarray}
 \hat{V}_{ee}^{N'}&=&\sum_{i=1}^{N'}\sum_{j>i}^{N'}\frac{1}{\vert\mathbf{r}_i-\mathbf{r}_j\vert},\label{eq:def_vee}\\
 \hat{V}^{N'}&=&\sum_{i=1}^{N'}v(\mathbf{r})\delta(\mathbf{r}_i-\mathbf{r}),\\
 \hat{T}^{N'}&=&-\frac{1}{2}\sum_{i=1}^{N'}\nabla^2_{\mathbf{r}_i},\\
 \hat{V}^{N'}_{emb}&=&\sum_{i=1}^{N'}v_{emb}(\mathbf{r})\delta(\mathbf{r}_i-\mathbf{r}).
 \end{eqnarray}
In FDET,  $\hat{V}^{N'}_{emb}$  is multiplicative.
This restriction  leads to specific  mathematical issues concerning the solutions of Eq. \ref{eq:emb_SE} and their relations to the exact solution of the full $N$-electron problem, i.e.  $E_v^{o}$ for energy and $\rho_v^{o}$ for density in the ground state. Some of them are discussed in the present work. 
 
FDET provides thus the interpretation of any equation of the form of Eq. \ref{eq:emb_SE}, in which the embedding potential depends on election densities,  as   the Euler-Lagrange equation for  $E^{\rho_2}_{v}[\rho_1]$,
in which $\Psi^{N'}$ is an auxiliary quantum descriptor for the $N'$-electron system used to represent the  density $\rho_1$ as:
     $\rho_1(\mathbf{r})=\langle\Psi^{N'}\vert\sum_{i=1}^{N'}\delta(\mathbf{r}_i-\mathbf{r})\vert\Psi^{N'}\rangle$.



FDET provides the unique relation (up to an additive constant) between the potential $v_{emb}(\mathbf{r})$ in Eq. \ref{eq:emb_SE}, the embedded wavefunction obtained in this equation, and $\rho_2(\mathbf{r})$. 
\begin{eqnarray}
    v_{emb}[\rho_1^{{Eq. \ref{eq:emb_SE}}}, \rho_2](\mathbf{r}) =  \int \frac{\rho_2(\mathbf{r}')}{|\mathbf{r}-\mathbf{r}'|} d\mathbf{r}' 
    + \left.\frac{\delta E_{xct}^{nad}[\rho, \rho_2]}{\delta \rho (\mathbf{r})} \right|_{\rho=\rho_1^{{Eq. \ref{eq:emb_SE}}}},\label{eq:embpotwf}
\end{eqnarray}
in which the second term is the functional derivative of the bi-functional $E_{xct}^{nad}[\rho_1, \rho_2]$ defined as:
 \begin{eqnarray}
E_{xct}^{nad}[\rho_1, \rho_2]\equiv E_{xct}[\rho_1+\rho_2]-E_{xct}[\rho_1]-E_{xct}[\rho_2],\label{def:Exctnad}
\end{eqnarray} 
where
\begin{eqnarray}
E_{xct}[\rho]\equiv F^{HK}[\rho]-\frac{1}{2}
\int \int 
\frac
{\rho(\mathbf{r})\rho(\mathbf{r}')}
{
\left|\mathbf{r}-\mathbf{r}\right|}
\mathrm{d}\mathbf{r}'
\mathrm{d}\mathbf{r},
\label{def:Exct}
\end{eqnarray}
and where $F^{HK}[\rho]$ is the universal Hohenberg-Kohn functional defined implicitly in the Levy's constrained search \cite{Levy1982}:
\begin{eqnarray}
F^{HK}[\rho]&\equiv&
\min_{\Psi\longrightarrow \rho}\left<\Psi\vert \hat{T}+\hat{V}_{ee}\vert \Psi\right> 
= \left<\Psi_{o}[\rho]\vert \hat{T}+\hat{V}_{ee} \vert \Psi_{o}[\rho]\right>\label{eq:def_TandVee}\\
&=& \left<\Psi_{o}[\rho]\vert \hat{T} \vert \Psi_{o}[\rho]\right>+ \left<\Psi_{o}[\rho]\vert \hat{V}_{ee} \vert \Psi_{o}[\rho]\right>=T[\rho]+V_{ee}[\rho]. \nonumber
\end{eqnarray}
~\\
%
As shown in Ref. \citenum{Wesolowski2008},  the minimiser defined in Eq. \ref{eq:constrminim}  and the density obtained from Eqs. \ref{eq:emb_SE},\ref{eq:embpotwf} are the same:
\begin{eqnarray}
\rho_1^{Eqs.\;\ref{eq:emb_SE},\ref{eq:embpotwf}}[\rho_2]=\rho_1^{o}[\rho_2], \label{eq:targetquestion}
\end{eqnarray}
for any $N-N'$-electron density $\rho_2$, provided $\rho_1^{o}[\rho_2]$ is $v$-representable\cite{Parr1989}. 

%
Only minima of  $E^{\rho_2}_{v}[\rho_1]$ can be obtained from the FDET eigenvalue equation.
It is possible that,  there exist such density  $\rho_2$ for which: 
\begin{eqnarray}
\inf_{\rho\rightarrow\forall_{\mathbf{r}}\big(\rho({\mathbf r})\ge \rho_2({\mathbf r}\big)} 
E^{HK}_v[\rho]=
E^{HK}_v[\rho_1^{lowest}[\rho_2]+\rho_2] \le E^{HK}_v[\rho_1^{Eqs.\;\ref{eq:emb_SE},\ref{eq:embpotwf}}[\rho_2]+\rho_2].\label{eq:min_inf_inequality}
\end{eqnarray}
If such density $\rho_2$ exists, the corresponding $\rho_1^{lowest}[\rho_2]$ (defined in Eq. \ref{eq:min_inf_inequality}), is an infimum  but not minimum of $E^{\rho_2}_{v}[\rho_1]$. It cannot be obtained from FDET therefore. 
%
This observation leads to questions addressed in the present work:
\begin{itemize}
\item Are there any densities  $\rho_2$,  for which $\rho_1^{lowest}[\rho_2]$ is not a minimum?
\item What is $\rho_1^{Eqs.\;\ref{eq:emb_SE},\ref{eq:embpotwf}}[\rho_2]$ in such a case?
\end{itemize}
%
%
Concerning the first issue, we identify a non-trivial class of  $N-N'$-electron densities $\rho_2$   (denoted with $\mathcal{A}$ below), for which that $\rho_1^{lowest}[\rho_2]$ cannot be obtained from Eq. \ref{eq:emb_SE}. It is shown by proving that the functional derivative of $E_{xct}^{nad}[\rho_1, \rho_2]$ with respect to $\rho_1$ does not exist for $\rho_1=\rho_v^{lowest}[\rho_2]$ and $\rho_2\in\mathcal{A}$ ({\it Lemma A}).
{\it Theorems I}-{\it II} provide inequalities for density and energy obtained from FDET eigenvalue equations for $\rho_2\in\mathcal{A}$. 
The derived  inequalities for energy and density are also given for the variant of FDET eigenvalue equation in which a non-interacting reference Hamiltonian is used instead of $V_{ee}$ in Eq. \ref{eq:emb_SE}. \\
%
%
~\\
{\it Definition:} class  $\mathcal{A}$\\

\begin{figure}
	\centering
	\includegraphics[scale=0.2,angle=90]{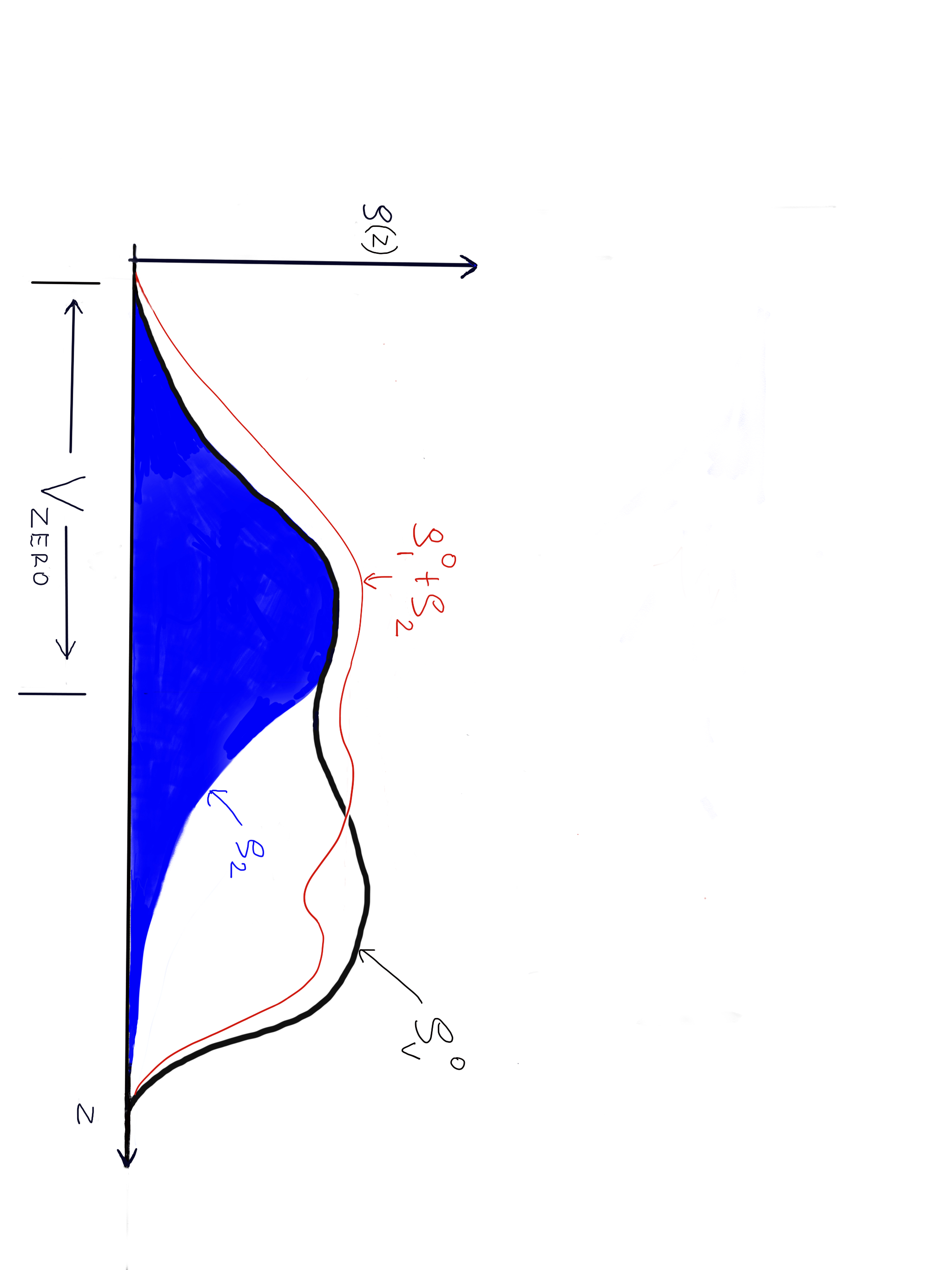}
	\caption{Symbolic representation of the densities considered in this work.  $\rho_2$ belonging to $\mathcal{A}$ is indicated with blue. The ground state density is indicated with  black. Red indicates the sum of $\rho_2$ and the stationary density obtained from Eq. \ref{eq:emb_SE}.  
	  }\label{fig:scheme}
\end{figure}

The $N-N'$-electron density 
$\rho_2(\mathbf{r})$ belongs to $\mathcal{A}$ 
if there exists a measurable volume element $V_{z}$ such that:
\begin{eqnarray}
\forall_{\mathbf{r}\in V_{z}}\rho_2(\mathbf{r})=\rho_v^{o}(\mathbf{r}), \label{def:setA1}
\end{eqnarray}
and 
\begin{eqnarray}
\forall_{\mathbf{r}}\rho_2(\mathbf{r})\le \rho_v^{o}(\mathbf{r}).\label{def:setA2}
\\ \nonumber
\end{eqnarray}
~\\
A density belonging to $\mathcal{A}$ is 
%
 shown schematically in Fig. \ref{fig:scheme}.\\

The pair of densities $\rho_2\in\mathcal{A}$ and the corresponding $\rho_1^{lowest}[\rho_2]$ represent a particular partitioning of the exact ground-state density $\rho_v^{o}$ as:
\begin{eqnarray}
\inf_{\rho\rightarrow\forall_{\mathbf{r}}\big(\rho({\mathbf r})\ge \rho_2({\mathbf r}\big)} 
E^{HK}_v[\rho]=
E^{HK}_v[\overbrace{\rho_1^{lowest}[\rho_2]}^{=\rho_v^o-\rho_2}+\rho_2]=E^{HK}_v[\rho_v^o]=E_v^o\;\mathrm{for}\;\rho_2\in\mathcal{A}.
\label{def:EatInfimum}
\end{eqnarray}
We point out  that FDET is formulated for a more general class of densities $\rho_2$ than the ones corresponding to some partitioning of $\rho_v^{o}$ which includes $\rho_2\in\mathcal{A}$ as a subset but includes also such $N-N'$-electron densities $\rho_2$ that  $\rho_2(\mathbf{r})> \rho_v^{o}(\mathbf{r})$ on some measurable volume.
\subsubsection*{FDET eigenvalue equation for  $\rho_2\in\mathcal{A}$}
$\rho_1^{Eqs.\;\ref{eq:emb_SE},\ref{eq:embpotwf}}[\rho_2]=\rho_1^{o}[\rho_2]$ only if the bi-functional  $v_{xct}^{nad}[\rho_1,\rho_2]$ exists for $\rho_1=\rho_1^{o}[\rho_2]$. 
The general condition 
\begin{eqnarray}
\forall_{\mathbf{r}}\rho_1[\rho_2]({\mathbf r})\ge 0
\end{eqnarray}
is necessary that the potential $v_{xct}^{nad}[\rho_1[\rho_2],\rho_2]$ exists. Obviously it is  automatically satisfied by densities obtained from Eqs.\;\ref{eq:emb_SE},\ref{eq:embpotwf} for any $\rho_2$.
Below, we show that a stronger condition for the existence of $v_{xct}^{nad}[\rho_1[\rho_2],\rho_2]$ applies  if  $\rho_2\in \mathcal{A}$.
\\
~\\
%
%
{\it 
\noindent{Lemma A:}}\\ 
The   potential $v_{xct}^{nad}[\rho_1,\rho_2]({\mathbf r})$ does not exist for   $\rho_1=\rho_v^{o}-\rho_2$  and  $\rho_2\in \mathcal{A}$, where $\rho_v^{o}$ is a $v$-representable ground-state electron density of $N$-electrons in a Coulombic potential $v$.
\\
~\\
{\it 
\noindent{Proof:}} 
The proof proceeds by {\it reductio ad absurdum}.  
Let assume the contrary, i.e.  the functional derivative $v_{xct}^{nad}[\rho_1,\rho_2]({\mathbf r})=\left.
\frac{\delta E_{xct}^{nad}[\rho_1,\rho_2]}{\delta \rho_1({\mathbf r})}
\right|_{\rho_1=\rho_v^{o}-\rho_2} $ exists. The  Gateaux differential 
$\delta E_{xct}^{nad}[\rho_1,\rho_2]$ is given by:
\begin{eqnarray}
\delta E_{xct}^{nad}[\rho_1,\rho_2]({\mathbf r})&=&\delta \big(E_{xct}[\rho_1+\rho_2]-E_{xct}[\rho_1]-E_{xct}[\rho_2]\big)=
{\delta E_{xct}[\rho_1+\rho_2]}
-
{\delta E_{xct}[\rho_1]}
\label{eq:prooflemma}
\\
\nonumber\\
&=& 
\lim_{h\rightarrow 0}
\frac{E_{xct}[\rho_1+h\cdot \Delta+\rho_2]-E_{xct}[\rho_1+ \rho_2]}{h}
- 
\lim_{h\rightarrow 0}
\frac{E_{xct}[\rho_1+h\cdot \Delta]-E_{xct}[\rho_1]}{h}.\nonumber
\end{eqnarray} 
The functional derivative exists only if the limit in the Gateaux differential does not depend on 
the function $\Delta({\mathbf r})$.  Below an example of $\Delta({\mathbf r})$ is provided, for which only the first limit  in Eq. \ref{eq:prooflemma} exists.

Concerning  the first Gateaux differential in Eq. \ref{eq:prooflemma}, 
($\lim_{h\rightarrow 0}
\frac{E_{xct}[\rho_1+h\cdot \Delta+\rho_2]-E_{xct}[\rho_1+ \rho_2]}{h}
$)
 it exists because $\rho_v^{o}$ is $v$-representable. 
The differential  is, therefore,  the same for any $\Delta({\mathbf r})$ such that  $\rho_1+h\cdot \Delta+\rho_2$ is admissible in $E_{xct}[\rho]$.
%
%
Let  us evaluate the Gateaux differential $\delta E_{xct}[\rho_1+\rho_2]$ using  $\Delta({\mathbf r})$ such that:\\
a) $\int \Delta({\mathbf r})d{\mathbf r}=0$;\\
b) $\Delta({\mathbf r})$ is bound from below and above;\\
c)  $\forall_{\mathbf{r}\in V_{z}}\Delta({\mathbf r})< 0$ and $\vert\Delta({\mathbf r})\vert\rightarrow 0$ faster than  $\rho_v^{o} ({\mathbf r})$ at $|{\mathbf r}|\rightarrow \infty$.\\
Conditions a)-c) are required to assure the that the density  $\rho_1+\rho_2+h\cdot\Delta$ is admissible in $E_{xct}[\rho]$:
a) is necessary for $\int \big(\rho_1({\mathbf r})+ \rho_2({\mathbf r})+h\cdot\Delta({\mathbf r})\big)d{\mathbf r}=N$;
b) assures that a sufficiently small $h$ can be found such that the sum $\rho_v^{o}({\mathbf r})+h\cdot \Delta({\mathbf r})>0$ at the maximal and minimal values of $\Delta({\mathbf r})$ (the assumption of $v$ being Coulombic  is used here - for Coulombic systems;  $\rho_v^{o}({\mathbf r})>0$ everywhere except at infinity);  c) guarantees that $\rho_v^{o}({\mathbf r})+h\cdot \Delta({\mathbf r})>0$ for ${\mathbf r}\rightarrow \infty$. 


Let us attempt to obtain the second differential  
($\lim_{h\rightarrow 0}\frac{E_{xct}[\rho_1+h\cdot \Delta]-E_{xct}[\rho_1]}{h}$) in Eq. \ref{eq:prooflemma}
using the same $\Delta({\mathbf r})$ and approach the limit also from the side of positive $h$.
%
For ${\mathbf r}\in V_{z}$, $\rho_1({\mathbf r})+h\cdot \Delta({\mathbf r})=h\cdot \Delta({\mathbf r})<0$, which makes it inadmissible in $E_{xct}[\rho]$. 
The limit $\lim_{h\rightarrow 0}\frac{E_{xct}[\rho_1+h\cdot \Delta]-E_{xct}[\rho_1]}{h}$ does not exist for this choice for $\Delta({\mathbf r})$ therefore.  As a result, neither the whole differential 
$\delta E_{xct}[\rho_1+\rho_2]$ nor he functional derivative $v_{xct}^{nad}[\rho_1,\rho_2]({\mathbf r})=\left.
\frac{\delta E_{xct}^{nad}[\rho,\rho_2]}{\delta \rho({\mathbf r})}
\right|_{\rho=\rho_1}$ exists for $\rho_2\in\mathcal{A}$. {\it This contradiction ends the proof.}\\
%
~\\
Following {\it Lemma A},  the potential 
$v_{xct}^{nad}[\rho_v^o-\rho_2,\rho_2]$ does not exist for 
$\rho_2\in \mathcal{A}$. The density $\rho_v^{o}-\rho_2$ cannot, therefore,  be obtained from  Eqs.\;\ref{eq:emb_SE},\ref{eq:embpotwf}.
 
 We note that the potential $v_{xct}^{nad}[\rho_1,\rho_2]$ depends on $\rho_1$. 
 Eqs. \ref{eq:emb_SE},\ref{eq:embpotwf} might have, therefore, another solution which, if exists, is a minimum. The density obtained form this solution will be denoted with $\rho_1^{Eqs.\;\ref{eq:emb_SE},\ref{eq:embpotwf}}$.
 
The proof of {\it Lemma A} makes it possible to identify important property of every $N'$-electron density obtained from  Eqs. \ref{eq:emb_SE},\ref{eq:embpotwf} for $\rho_2\in \mathcal{A}$.\\
~\\
{\it
\noindent{Theorem I:}}\\
For Coulombic external potential $v$ and $\rho_2\in \mathcal{A}$, 
there exists a measurable volume element contained in $V_z$ ($V'_z\subset V_z$), such that: 
\begin{eqnarray}
\forall_{\mathbf{r}\in V'_z}
\rho_1^{Eqs.\;\ref{eq:emb_SE},\ref{eq:embpotwf}}[\rho_2]({\mathbf r})&>&0.
\label{TheoremI}
\end{eqnarray}
{\it
\noindent{Proof:}}
Assuming the contrary, i.e., $\forall_{\mathbf{r}\in V_z}\rho_1^{Eqs.\;\ref{eq:emb_SE},\ref{eq:embpotwf}}[\rho_2]({\mathbf r})=0$, leads to contradiction.
According to {\it Lemma A},  $v_{xct}^{nad}[\rho_1,\rho_2]$ does not exist if $\forall_{\mathbf{r}\in V_z}\rho_1({\mathbf r})=0$. The density $\rho_1^{Eqs.\;\ref{eq:emb_SE},\ref{eq:embpotwf}}[\rho_2]$ cannot, therefore, be obtained from Eqs. \ref{eq:emb_SE},\ref{eq:embpotwf}. {\it End of the proof}.\\

The inequality for density given in Eq.  \ref{TheoremI} leads to the inequality for energy.\\ 
~\\
{\it
\noindent{Theorem II}}\\
~\\
For Coulombic external potential $v$ and $\rho_2\in \mathcal{A}$,
\begin{eqnarray}
E^{HK}_{v}\big[\rho_1^{Eqs.\;\ref{eq:emb_SE},\ref{eq:embpotwf}}[\rho_2]+\rho_2\big]
>E^{HK}_{v}\big[\rho_v^{o}\big]=E^o_v. 
\label{TheoremII}\\ \nonumber
\end{eqnarray}
{\it Proof:} The sharp inequality follows from $\rho_1^{Eqs.\;\ref{eq:emb_SE},\ref{eq:embpotwf}}[\rho_2]+\rho_2\neq \rho_v^{o}$ and the Hohenberg-Kohn theorems. {\it End of the proof.}\\
~\\
{\it Lemma A}, {\it Theorem I}, and {\it Theorem II}  concern  interacting Hamiltonians used in Eq. \ref{eq:emb_SE}. They analogues can be formulated in a straightforward manner  for the variant of FDET,  in which the non-interacting reference $N'$-electron Hamiltonian is used.  In this case, the embedded wavefunction has  the form of a single determinant $\Phi^{N'}$, and the FDET eigenvalue equation corresponding to Eqs. \ref{eq:emb_SE} and 
\ref{eq:embpotwf} reads: 
\begin{eqnarray}
    \left( \hat{T}^{N'}+\hat{V}^{N'}+\hat{V}_{J}^{N'}[\rho_1^{Eq. \ref{eq:KSCED0}}]+\hat{V}^{N'}_{xc}[\rho_1^{Eq. \ref{eq:KSCED0}}+\rho_2]
    + \hat{V}_{J}[\rho_2]+\hat{V}_{t}^{nad}[\rho_1^{Eq. \ref{eq:KSCED0}},\rho_2]    \right) \Phi^{N'} = \lambda^{N'}\cdot\Phi^{N'}, \label{eq:KSCED0} 
 \end{eqnarray}
 where
 \begin{eqnarray}
   \rho_1^{Eq. \ref{eq:KSCED0}}(\mathbf{r})&=&\langle\Phi^{\mathrm{Eq.}\;\ref{eq:KSCED0}}\vert\sum_{i=1}^{N'}\delta(\mathbf{r}_i-\mathbf{r})\vert\Phi^{\mathrm{Eq.}\;\ref{eq:KSCED0}}\rangle,\\
   v_{J}[\rho](\mathbf{r})&=&\int \frac{\rho(\mathbf{r})}{|\mathbf{r}-\mathbf{r}'|} d\mathbf{r}',
 \end{eqnarray}
 $\hat{V}^{N'}_{xc}[\rho]$ is  the conventional exchange-correlation potential defined in Kohn-Sham formulation \cite{Kohn1965} of density functional theory, which is the functional of the $N$-electron density,
 whereas
 the $v_{t}^{nad}[\rho_1,\rho_2](\mathbf{r})$  component of the FDET embedding potential 
 is given by  the bi-functional:
  \begin{eqnarray}
 v_{t}^{nad}[\rho_1,\rho_2](\mathbf{r})&=&
  \left.\frac{\delta T_{s}^{nad}[\rho, \rho_2]}{\delta \rho (\mathbf{r})} \right|_{\rho=\rho_1},
 \end{eqnarray}
where
 $T_{s}^{nad}[\rho, \rho_2]$ is the kinetic component of $E_{xct}^{nad}[\rho_1, \rho_2]$:
 \begin{eqnarray}
T_{s}^{nad}[\rho_1, \rho_2]&\equiv& T_{s}[\rho_1+\rho_2]-T_{s}[\rho_1]-T_{s}[\rho_2]\label{def_tnad}\\
&=&
\min_{\Phi^N\longrightarrow \rho_1+\rho_2}\left<\Phi^N\vert \hat{T}\vert \Phi^N\right>-
\min_{\Phi^{N'}\longrightarrow \rho_1}\left<\Phi^{N'}\vert \hat{T}\vert \Phi^{N'}\right>-
\min_{\Phi^{N-N'}\longrightarrow \rho_2}\left<\Phi^{N-N'}\vert \hat{T}\vert \Phi^{N-N'}\right>.
\nonumber
\end{eqnarray}

For this variant of FDET, the analogue of {\it Lemma A} can be formulated.\\
~\\
{\it 
\noindent{Lemma B:}}\\ 
The   potential $v_{t}^{nad}[\rho_1,\rho_2]({\mathbf r})$ does not exist for   $\rho_1=\rho_v^{o}-\rho_2$  and  $\rho_2\in \mathcal{A}$, where $\rho_v^{o}$ is a $v_s$-representable ground-state electron density of $N$-electrons in a Coulombic potential $v$.\\
The proof  of  {\it Lemma B} uses the same steps as the ones for {\it Lemma A} applied  not for  $E_{xct}^{nad}[\rho_1,\rho_2]$ but for $T_{s}^{nad}[\rho_1,\rho_2]$.
They will  not be given here to avoid redundant text. It is, however, worthwhile to point out that  this variant of FDET concerns only such total systems for which $\rho_v^{o}$ is $v_s$-representable.

Moreover, as shown in Ref. \citenum{Wesolowski2008},  the minimiser defined in Eq. \ref{eq:constrminim}  and the density obtained from Eq. \ref{eq:KSCED0} are the same
 for any $N-N'$-electron density $\rho_2$, provided $\rho_1^{o}[\rho_2]$ and $\rho_v^{o}$ are both $v_s$-representable\cite{Parr1989}. 

The inequalities corresponding to Eqs. \ref{TheoremI} and \ref{TheoremII}
read: 
%
\begin{eqnarray}
\forall_{\mathbf{r}\in V'_z}
\rho_1^{Eq.\;\ref{eq:KSCED0}}[\rho_2]({\mathbf r})&>&0
\;\mathrm{for}\;\rho_2\in\mathcal{A}, 
\label{TheoremIKS}
\end{eqnarray}
\begin{eqnarray}
E^{HK}_{v}\big[\rho_1^{Eq.\;\ref{eq:KSCED0}}[\rho_2]+\rho_2\big]
>E^{HK}_{v}\big[\rho_v^{o}\big]=E^o_v \;\mathrm{for}\;\rho_2\in\mathcal{A}.  \label{TheoremIIKS}\\ \nonumber
\end{eqnarray}
%

%


\subsubsection*{Discussion}
The principal result of this work is showing that  the total electron density represented as a sum of $\rho_2$ and a component given as the solution of the corresponding exact FDET eigenvalue equation, cannot be equal to the exact total ground-state electron density $\rho_v^{o}$ if  the densities 
$\rho_2(\mathbf{r})$ are equal $\rho_v^{o}(\mathbf{r})$ on some measurable volume (Eqs. \ref{TheoremI} and \ref{TheoremIKS}).
This result might seem counterintuitive and originates from the fact that the FDET embedding potential is given as the functional derivative of a functional depending on the embedded $N'$-electron density. 
%
%
Even a "lucky choice" of $\rho_2$ made in a FDET based multi-scale simulation method, 
such that $\rho_2$ coincides with the exact ground state density in some domains in 3D,
leads to the total density which is different than the exact one.  
As a result, the total energy given by the Hohenberg-Kohn functional lies above $E_v^{o}$ in case of exact functionals used in FDET functional for energy and embedding potential.
The condition $\forall_{\mathbf{r}}\rho_2(\mathbf{r})\le \rho_v^{o}(\mathbf{r})$ does not guarantee reaching the exact total energy and density. A stronger condition -  $\forall_{\mathbf{r}}\rho_2(\mathbf{r})< \rho_v^{o}(\mathbf{r})$ - is needed. 

The  exact solution of the FDET eigenvalue equation applied for  $\rho_2$, which 
is piecewise equal to the exact density, was to our knowledge, not addressed directly  in the literature so far. 
The equality of the $\rho_2+\rho_1^{o}=\rho_v^{0}$ for $\rho_2\in\mathcal{A}$ for such $\rho_2$ that $\forall_{\mathbf{r}}\rho_2(\mathbf{r})\le\rho_v^{o}(\mathbf{r})$  is rather a silent assumption made commonly. 
Ref.  \citenum{Gritsenko2013} is a notable exception, in which this issue was mentioned.
The variant of FDET eigenvalue equation given here in Eq. \ref{eq:KSCED0} used in subsystem approach to the Hohenberg-Kohn theorems formulated by  Cortona \cite{Cortona1991},  was considered as a tool to get the exact total ground-state density $\rho_v^{o}$, i.e. in a narrower scope than  it is used in FDET. 
We recall  at this point that  FDET admits in this equation the densities  $\rho_2$ belonging to a larger set for which $\rho_1^{Eq. \ref{eq:KSCED0}}+\rho_2\ne \rho_v^{o}$. 
Even  $N-N'$-electron densities $\rho_2$,  such that  $\rho_2(\mathbf{r})>\rho_v^{o}(\mathbf{r})$ are admissible in FDET although the  total density corresponding to either the minimum or infimum obtained from Eq. \ref{eq:constrminim} or Eq. \ref{eq:min_inf_inequality} differs from $\rho_v^{o}$.
In Ref.  \citenum{Gritsenko2013},  it was noted that the density $\rho_1^{target}=\rho_v^{0}-\rho_2$ may not be obtainable  from Eq. \ref{eq:KSCED0} if  $\rho_2\in\mathcal{A}$. 
The considerations presented in Ref. \citenum{Gritsenko2013}  excluded, therefore, such a case explicitly. The present work shows  that $\rho_1^{Eq. \ref{eq:KSCED0}}+\rho_2\ne \rho_v^{o}$  is not only a  possibility but is always the case if  $\rho_2\in\mathcal{A}$. FDET eigenvalue equation (Eq. \ref{eq:KSCED0}) possesses, nevertheless, a solution for which $\rho_1^{Eq. \ref{eq:KSCED0}}+\rho_2\ne \rho_v^{o}$.

The present work exposes also  a fundamental difference between the embedding methods using only electron density $\rho_2$ as the descriptor for $N-N'$-electrons, for which FDET provided the exact formal framework,  and the method using finer descriptors. Let us consider the case of the pseudopoptential theory by   Phillips-Kleinman  \cite{Phillips1959} as an example. The in the pseudo-Schr\"odinger $N'$-electron equation the embedding operator is expressed by means of orbitals used for $N-N'$ electrons not taken explicitly in the $N'$-electron Hamiltonian. If the orbitals are such that the corresponding $N-N'$-electron density is piecewise equal to the exact $N$-electron density ($\rho_2\in\mathcal{A})$), the Phillips-Kleinman eigenvalue equation 
would still lead to the exact total electron density by construction, whereas  FDET wouldn't.
This qualitative difference, is due to the fact that the exact pseudopotential is not a multiplicative operator in contrast to the FDET embedding potential which is given as a functional derivative. The existence of such a derivative for an arbitrary $\rho_2$ is the issue specific for FDET. A more detailed analysis of the relation between the FDET embedding potential and the exact pseudopotential is given in Ref. \citenum{Wesolowski2013}.
%
%

%
%
The present work relates the  relation between $v_t^{nad}[\rho_1,\rho_2]({\mathbf r})$ and the functional $v_s[\rho]$ defined for $v_s$-representable densities \cite{Parr1989}.
\begin{eqnarray}
v_t^{nad}[\rho_1,\rho_2]({\mathbf r})=v_s[\rho_1]({\mathbf r})-v_s[\rho_1+\rho_2]({\mathbf r}) +\mathrm{const}.\label{eq:vtnadinv0}
\end{eqnarray}
The above relation has been discussed in the literature and used in practice 
for  arbitrarily chosen pairs $\rho_1$ and $\rho_2$, in order to 
 generate reference embedding potentials for the sake of development of approximantions  for $v_t^{nad}[\rho_1,\rho_2]$ or as a way to avoid the use of explicit approximants  in practical simulations
\cite{Roncero2008,Wesolowski2009,Goodpaster2010,Fux2010org,deSilva2012b,artiukhinExcitationEnergiesFrozendensity2015,Nafziger2017,Banafsheh2018,Schnieders2018,Banafsheh2022}. 
In these approaches, Eq. \ref{eq:vtnadinv0} is used in the procedure consisting of the following steps:
\begin{itemize}
\item[a)] the density 
 $\rho_{Tot}$ is generated using some Kohn-Sham based method, which provides the total Kohn-Sham potential equal to $v_s[\rho_{Tot}](\mathbf{r})$; 
\item[ b)] the density $\rho_{Tot}$
 is partitioned as using some arbitrary criteria $\rho_{Tot}=\rho_1^{part}+\rho_2^{part}$;
 \item[c)] the potential $v_s[\rho_1^{part}](\mathbf{r})$ is generated using the procedure to "invert" the Kohn-Sham equation, i.e. to obtain the potential $v_s[\rho]$ corresponding to a given $\rho$. It is possible analytically  if $\int \rho_1^{part}](\mathbf{r})d\mathbf{r}=1$ or if $\int \rho_1^{part}](\mathbf{r})d\mathbf{r}=2$ \cite{Wesolowski2009,deSilva2012b} or made numerically (see the reviews  in Refs. \citenum{Banafsheh2018,Schnieders2018});
 \item[d)] the potential $v_t^{nad}[\rho_1^{part},\rho_2^{part}]$ is obtained form Eq. \ref{eq:vtnadinv0}. 
 \end{itemize} 
According to {\it Lemma B}, the above procedure procedure can be numerically sound, only if $\rho_{Tot}$ is  partitioned assuring that  $\rho_2^{part}\notin \mathcal{A}$.
 {\it Lemma B}  completes also  the analysis of Eq. \ref{eq:vtnadinv0} presented in Ref. \citenum{Wesolowski2022}, where it was indicated that the potential $v_t^{nad}[\rho_1^{part},\rho_2^{part}]$ 
might not be unique  if $\rho_2^{part}\in\mathcal{A}$.  {\it Lemma B} shows that such potential does not exist.
The relation given in Eq. \ref{eq:vtnadinv0} can be used  also for the intuitive justification 
that the potential $v_t^{nad}[\rho_v^{o}-\rho_2,\rho_2]$ does not exist 
{\it Lemma B}  if $\rho_2\in\mathcal{A}$. 
Kohn-Sham equation can lead to 
such $\rho_1$ that disappear on $V_z$ only if the total effective potential is infinite at  $\mathbf{r}\in V_z$.
The author is grateful to Prof. T. Gould for his comments on the general idea presented in this work and to Dr. Pierre-Olivier Roy for his comments on the earlier version of this manuscript.
{
}
\bibliography{250506_biball_forTheorem}

\end{document}